# Octahedral tilting induced ferroelectricity in $ASnO_3/BSnO_3$ superlattice

Hyunsu Sim[1], S. W. Cheong[2], and Bog G. Kim[1,*]

[1] *Department of Physics, Pusan National University, Pusan, 609-735, South Korea*
[2] *Rutgers Center for Emergent Materials and Department of Physics and Astronomy, Rutgers University, Piscataway, NJ 08854, USA.*

The effect of octahedral tilting of $ASnO_3$ (A = Ca, Sr, Ba) parent compound and bi-color $ASnO_3/BSnO_3$ superlattice (A, B = Ca, Sr, Ba) is predicted from density-functional theory. In $ASnO_3$ parent compound, the structural phase transition as a function of A-site cation size is correlated with magnitude of the two octahedral tilting modes ($a^-a^-c^0$ tilting and $a^0a^0c^+$ tilting). The magnitude of octahedral tilting modes in the superlattices is analyzed quantitatively and is associated with that of constituent parent materials. $ASnO_3/BSnO_3$ superlattices show the hybrid improper ferroelectricity resulting from the coupling of two octahedral tilting modes ($a^-a^-c^0$ tilting and $a^0a^0c^+$ tilting), which are also responsible for the structural phase transition from the tetragonal phase to the orthorhombic phase. Ferroelectricity due to A-site mirror symmetry breaking is secondary order parameter for the orthorhombic phase transition in the bi-color superlattice and is related with $\Gamma_{5-}$ symmetry mode. The coupling between tilting modes and ferroelectric mode in the bi-color superlattice of $ASnO_3/BSnO_3$ is analyzed by group theory and symmetry mode analysis.

Recently, there has been considerable interest in transparent conducting oxide (TCO) which exhibit optical transparency with high electrical conductivity not only for technological importance as touch screen and dye-synthesized solar cell but also scientific importance [1-4]. $ASnO_3$ is one of potential candidates for next generation TCO materials with perovskite structure [5-22]. In this material, Sn ion forms an octahedron with 6 neighboring oxygen atoms and $SnO_6$ octahedral are linked in three dimensional networks with corner sharing oxygen. $ASnO_3$ shows various structural forms depending on the size of A site atoms also depending on the external parameters, such as temperature and pressure.

The most common structure of $ASnO_3$ is orthorhombic phase, in which octahedral tilting is important ingredient [23-26]. Three dimensional network of octahedral tilting in $ASnO_3$ can be characterized by so called Glazer notation [23]. The Glazer notation is describes the octahedral tilting using symbol $a^\# b^\# c^\#$, in which the literals refer to tilt around the [100], [010], and [001] directions of the cubic perovskite, and the superscript # takes the value 0, +, or – to indicate no tilt or tilts of successive octahedral in the same or opposite sense. The orthorhombic structure of $ASnO_3$ can be classified by $a^- a^- c^+$ tilting [6-8, 23]. This tilting is basically originated from the interatomic forces and the size of A site atoms. When A site atoms are larger for $BaSnO_3$, the tilting does not occur and the $BaSnO_3$ belongs to cubic perovskite [5, 22]. Whereas smaller size of A site atoms for Sr, Ca, the tilting occurs to maintain the environment of the B cation unchanged. The physical properties are closely associated with the amount of tilting since the overlapping integral of $BO_6$ octahedral is one of main sources of various electronic properties [8, 10-13, 21]. The electronic properties such as band gap, density of state, and effective carrier mass are also closely related with octahedral tilting too [5-7, 21, 22].

The octahedral tilting can also be important for various perovskite derived systems [27-30]. Bousquet *et al.* [27] show that by layering perovskites in an artificial superlattice $SrTiO_3/PbTiO_3$, a polarization can arise from the coupling of two octahedral tilting modes. N. A. Benedek and C. J. Fennie [28] also introduced hybrid improper ferroelectricity concept in $(ABO_3)_2(AO)$ layered perovskites. They demonstrate that the polarization, P, arises from a rotation pattern (tilting of octahedral) that is a combination of two nonpolar lattice modes with different symmetries. J. M. Rondinelli and C. J. Fennie [29] also find that octahedral rotation induced ferroelectricity in cation ordered perovskites, $(ABO_3)/(A'B'O_3)$ systems.

Here, we report the detailed analysis of the octahedral tilting of $ASnO_3$ parent compounds and $ASnO_3/BSnO_3$ superlattices. We optimized the structure by using first principle calculation. Then, the quantitative analysis has been applied to the octahedral tilting of the each system. The quantitative tilting analysis is presented for the parent compounds. Next, $ASnO_3/BSnO_3$ superlattices indeed show the hybrid improper ferroelectricity resulting from the coupling of two octahedral tilting modes. The primary and secondary order parameters for the structural phase transition are identified.

We performed the first principles calculations with the local density approximation (LDA) [31] to the density functional theory and the projector-augmented-wave method as implemented in VASP [32, 33]. We considered the following valence electron configuration: $3p^64s^2$ for Ca, $4s^24p^65s^2$ for Sr, $5s^25p^66s^2$ for Ba, $4d^{10}5s^25p^2$ for Sn, and $2s^22p^4$ for Oxygen.

Electronic wave functions are expanded with plane waves up to a kinetic-energy cutoff of 400 eV except for structural optimization, where kinetic energy cutoff of 520 eV has been applied in order to reduce the effect of Pulray stress. Momentum space integration is performed using $4 \times 5 \times 5$ Monkhorst-Pack k-point mesh [34]. With the given symmetry of perovskite imposed, lattice constants and internal coordinates were fully optimized until the residual Hellmann-Feyman forces became smaller that $10^{-3}$ eV/Å. Calculating total energy as a function of fixed volume is used to obtain the equation of states, and at each given volume, internal atomic coordinate was fully optimized. In order to get phonon dispersion curve and phonon partial density of state, frozen phonon calculation has been applied on a $2 \times 2 \times 2$ supercell (containing 160 atoms in superlattice configuration) using phonopy program [35]. Spontaneous polarization was obtained by using Berry phase technique [36]. ISOTROPY and AMPLIMODES program were utilized to check group subgroup relationship and to quantify octahedral tilting [37, 38].

The oxygen octahedral tilting can induce space group change as shown in Fig. 1. Without any tilting, the parent compound of $ASnO_3$ is $Pm\bar{3}m$ cubic structure (Fig. 1(a)). Depending on the tilting instability, two different subgroups are illustrated. With $a^0a^0c^-$ tilting, the $I4/mcm$ tetragonal phase can be formed (Fig. 1(b)) and with more complicated $a^-a^-c^+$ tilting, $Pnma$ orthorhombic phase can be produced (Fig. 1(c)) [23, 26, 29]. It is important to mention that both phases are non-polar space group. The Total energy vs. volume with given space group for one perovskite formula unit cell is calculated to show equation of state diagram [39]. In $CaSnO_3$, lowest energy state is orthorhombic phase (Fig. 1(d)). The octahedral tilting possibility is checked for $BaSnO_3$, $SrSnO_3$, and $CaSnO_3$. The results are summarized in Fig. 1(e) for total energy vs space group symmetry. For Ca compound, the small ionic size of Ca is related with small tolerance factor and the orthorhombic phase has ~ 0.45 eV/f. u. lower energy than the tetragonal phase. For Sr compound, the orthorhombic phase has ~ 0.06 eV/f. u. lower energy than the tetragonal phase. For Ba compound, because the ionic size of Ba is largest among three compounds, the orthorhombic phase has nearly same energy with tetragonal and cubic phase within our calculation (see suppl. Fig. S1(a) and Fig. S1(b) for detailed calculation [40]). Tolerance factor is closely related with tilting angle in orthorhombic phase and the two different tilting angles ($a^-$ and $c^+$) are analyzed for $BaSnO_3$, $SrSnO_3$, and $CaSnO_3$ in Pnma phase. As ionic size decreases, the titling angle becomes larger shown in Fig. 1(f). One can conclude that the octahedral tilting instability is closely related with inverse of A-site cation size in parent $ASnO_3$ compound.

Now let us turn our attention to bi-color superlattice [27, 29] of $ASnO_3/BSnO_3$. Without any tilting, the space group of superlattice of $ASnO_3/BSnO_3$ is $P4/mmm$ tetragonal (Fig. 2(a)). Blue ball in Fig. 2(a) represents larger cation (A) and orange ball in Fig. 2(a) represents small cation (B). Even without tilting, c-axis can be two times larger than that of the cubic perovskite and thus the tetragonal space group must be assigned for $ASnO_3/BSnO_3$ superlattice. Group theory analysis shows that $a^-a^-c^+$ octahedral tilting stabilize $Pmc2_1$ orthorhombic phase in superlattice, shown in Fig. 2(b) [26, 29, 37]. Now unit cell contains 4 formula units of perovskite (20 atoms) and $Pmc2_1$ space group is actually polar space group [29, 37]. We have checked the equation of state for all possible superlattice combination of parent compounds, namely $BaSnO_3/SrSnO_3$, $BaSnO_3/CaSnO_3$, and $SrSnO_3/CaSnO_3$ (shown in Fig. 2(c), 2(d), and 2(e)).

First of all, the polar orthorhombic phase ($Pmc2_1$) is always more stable than the tetragonal phase ($P4/mmm$). Note that experimental ground states for parent compound are orthorhombic for $SrSnO_3$ and $CaSnO_3$ and cubic for $BaSnO_3$ [5, 7, 10, 22]. By forming superlattice of cubic $BaSnO_3$ and orthorhombic $BSnO_3$ (B = Ca, Sr), the octahedral tilting instability of orthorhombic parent compound plays important role. Secondly, the energy difference of orthorhombic and tetragonal phase is closely related with tilting instability of constituent materials, smallest for $BaSnO_3/SrSnO_3$ and largest for $SrSnO_3/CaSnO_3$. It can be said that average tolerance factor of two different cations can be important parameter for the orthorhombic phase stability. Thirdly, stable volume of the orthorhombic phase is always smaller than that of the tetragonal phase. The minimum points in the orthorhombic phase locate in the left side of the tetragonal minimum points. This can be also easily understood since the octahedral tilting is related with volume shrink in perovskite structure [23, 25]. Fourthly, the octahedral tilting angles are closely correlated to the structural parameters. The octahedral tilting in superlattices is analyzed for three different superlattices and the tilting angles ($a^-$ and $c^+$) are depicted in Fig. 2(f). Two dashed lines are the mean value of tilting angles of parent compounds (see Fig. 1(f)). In the first order, the tilting angles of superlattice correlated with the average tilting angles of the parent compounds. Exact values of tilting angles can be complicated function of strain energy of superlattice and can be obtained by theory or/and experiment.

In order to check the phase stability of the tetragonal and orthorhombic phases, phonon dispersion curves are obtained for superlattice structures. The calculations of the dynamic properties were performed using the force constant method [35]. In order to investigate phonon dispersion curve, a $2 \times 2 \times 2$ supercell (containing 80, 160 atoms in superlattice configuration) of the tetragonal and orthorhombic cell was used. The force constants were calculated for the displacement of atoms of up to 0.04 Å and the dynamical matrix at each q point of Brillouin zone was constructed by Fourier transforming the force constant calculated at the Γ point and the zone boundaries. Fig. 3(a), 3(b), and 3(c) are the phonon dispersion curves for $BaSnO_3/SrSnO_3$, $BaSnO_3/CaSnO_3$, and $SrSnO_3/CaSnO_3$ in the tetragonal phase. The negative value of phonon frequencies represents the imaginary unstable phonon mode. For $BaSnO_3/SrSnO_3$ superlattice, the phonon spectrum shows unstable mode near X, M and A Brillouin point. For $BaSnO_3/CaSnO_3$, and $SrSnO_3/CaSnO_3$ superlattice, imaginary phonon mode appears nearly all the Brillouin zones including zone center Γ point. Fig. 3(d), 3(e), and 3(f) depict the phonon dispersion curves for BaSnO$_3$/SrSnO$_3$, BaSnO$_3$/CaSnO$_3$, and SrSnO$_3$/CaSnO$_3$ in the orthorhombic phase. It is clear that phonon mode does not show any unstable mode in the orthorhombic phase. This result indicates that the tetragonal phase of superlattice is unstable toward the orthorhombic phase, meaning spontaneous transformation from the tetragonal phase to the orthorhombic phase occur.

Figure 4(a), 4(b), and 4(c) depict the electronic structures (band structures and density of the states) of BaSnO$_3$/SrSnO$_3$, BaSnO$_3$/CaSnO$_3$, and SrSnO$_3$/CaSnO$_3$ in the orthorhombic phase. The band structures of three superlattices are quite similar in shape. Valence band is quite flat throughout the Brillouin zone and the valence band maximum (VBM) is located in S point (0.5, 0.5, 0.000). The conduction bands show strong dispersion along Brillouin zone and the conduction band minimum (CBM) occur in zone center Γ point. It is also interesting to indicate that such a large dispersion near the CBM will results in small effective mass of conduction electron in three superlattices. These band structure shapes are quite similar with that of parent compound in the orthorhombic phase (see Suppl. Fig. S2 [40]). The cubic BaSnO$_3$ has different band shape due to Brillouin zone assignment (see Suppl. Figs S2(a) and S2(b) [40]). The band gaps of the three superlattices are 1.739, 1.995, and 2.470 eV for BaSnO$_3$/SrSnO$_3$, BaSnO$_3$/CaSnO$_3$, and SrSnO$_3$/CaSnO$_3$. It indicate that the size of octahedral distortion correlates with the size of band gap. We have calculated the spontaneous polarization of the superlattices (shown in Fig. 4(d)) by using berry phase calculation (see Suppl. Fig. S3 [40]). The coupling between octahedral tilting and A site cation displacements in opposite directions will results in the spontaneous polarization in the bi-color superlattice of the orthorhombic phase. The magnitude of polarization is smallest for BaSnO$_3$/SrSnO$_3$ and largest for BrSnO$_3$/CaSnO$_3$.

We are now in a position to discuss space group change and ferroelectric properties of the superlattice with quantitative symmetry mode analysis [37, 38]. In order to analyze the spontaneous phase transition from tetragonal to orthorhombic phase in the superlattice, we have used ISOTROPY for coupled mode analysis [37], AMPLIMODES for quantitative symmetry mode analysis [38]. Having the crystallographic information of two phases optimized by first principle calculation (the lattice constant and other properties are summarized in table I and atomic positions are given in supplementary Table I. [40]), the final results are summarized in Table II (see supplementary Table II data for more detailed calculation, [40]). The structural transition from tetragonal to orthorhombic phase is associated with three main modes with large amplitude, $a^0a^0c^+$ tilting, $a^-a^-c^0$ tilting, and opposite displacement of two A-site cations. In BaSnO$_3$/SrSnO$_3$ system, the symmetry of each mode is $M_{3+}$, $M_{5-}$ and $\Gamma_{5-}$ [38, 40]. The k-vectors of $M_{3+}$ and $M_{5-}$ mode are (1/2, 1/2, 0), which is related with cell doubling in *a-b* plane. These two modes are primary order parameter for the structural phase transition. In other words, the coupled mode analysis shows that non-zero amplitude of $M_{3+}$ and $M_{5-}$ mode will results in space group change from P*4/mmm* to P*mc2$_1$*. The $\Gamma_{5-}$ mode is addition mode in bi-color superlattice and the symmetry is already broken due to different two A-site atoms. In the structural phase transtion, the non-zero amplitude of $\Gamma_{5-}$ mode act as secondary order parameter responsible for spontaneous polarization [40]. For $BaSnO_3/CaSnO_3$ and $SrSnO_3/CaSnO_3$ system, the symmetry of two tilting mode are $M_{2+}$ and $M_{5-}$. The amplitudes of tilting modes, of course, are directly associated with the tilting angles in orthorhombic phase (Fig. 2(f)). Ferroelectric polarization is related with amplitude of $\Gamma_{5-}$ as well as the size difference of A-site cation. For the similar size mismatch of A-site cation, as the amplitude of $\Gamma_{5-}$ increases, the value of polarization increases (Ba/Sr vs Sr/Ca). However, due to strong cation size mismatch dependence of ferroelectric polarization, the linear relationship between the amplitude of $\Gamma_{5-}$ and the size of polarization does not hold (Ba/Ca) [41].

In conclusion, we have presented the detailed octahedral tilting analysis in $ASnO_3$ parent compound and $ASnO_3/BSnO_3$ superlattices based on first principle pseudopotential calculation. The octahedral tilting mode is closely associated with structural phase transition in the $ASnO_3$ parent compound and related with the size of A-site cation. Two octahedral tilting modes in superlattice are responsible for structural phase transition from the tetragonal phase to the orthorhombic phase. The magnitude of octahedral tilting modes in the superlattices is analyzed quantitatively and is associated with that of constituent parent materials. Ferroelectricity due to A-site mirror symmetry breaking is secondary order parameter for the orthorhombic phase transition in the bi-color superlattice and is related with $\Gamma_{5-}$ symmetry mode. The coupling between tilting modes and ferroelectric mode in the bi-color superlattice of $ASnO_3/BSnO_3$ open new engineering possibility of $SnO_6$ octaheral contating perovskite.

We acknowledge Prof. James Rondinelli for his help in polarization analysis. This work was supported by NSF of Korea (KRF-2012-0000964). Computational resources have been provided by KISTI Supercomputing Center (Project No. KSC-2012-C1-15).

**References**

[1] D. S. Ginley and C. Bright, Mater. Res. Bull. **25**, 15 (2000).

[2] G. Thomas, Nature **389**, 907 (1997).

[3] H. Hosono, Thin Solid Films **515**, 6000 (2007).

[4] C. G. Granqvist and A. Hultåker, Thin Solid Films **411**, 1 (2002).

[5] D. J. Singh, D. A. Papaconstantopoulos, J. P. Julien, and F. Cyrot-Lackmann, Phys. Rev. B **44**, 9519 (1991).

[6] H. Mizoguchi, P. M. Woodward, S. H. Byeon, and J. B. Parise, J. Am. Chem. Soc. **126**, 3175 (2004).

[7] H. Mizoguchi, H. W. Eng, and P. M. Woodward, Inorg. Chem. **43**, 1667 (2004).

[8] H. Mizoguchi, P. M. Woodward, C. -H. Park, and D. A. Keszler, J. Am. Chem. Soc. **126**, 9796 (2004).

[9] H. F. Wang, Q. Z. Liu, F. Chen, G. Y. Gao, W. Wu, and X. H. Chen, J. Appl. Phys. **101**, 106105 (2007).

[10] W. Zhang and J. Tang, J. Mater. Res. **3**, 1 (2007).

[11] M. Yoshida, T. Katsumata, and Y. Inaguma, Inorg. Chem. **47**, 6296 (2008).

[12] X. Luo, Y. S. Oh, A. Sirenko, P. Gao, T. A. Tyson, K. Char, and S. -W. Cheong, Appl. Phys. Lett. **100**, 172112 (2012).

[13] H. J. Kim, U. Kim, H. M. Kim, T. H. Kim, H. S. Mun, B. -G. Jeon, K. T. Hong, W. -J. Lee, C. Ju, K. H. Kim, and K. Char, Appl. Phys. Express **5**, 061102 (2012).

[14] S. S. Shin, J. S. Kim, J. H. Suk, K. D. Lee, D. W. Kim, J. H. Park, I. S. Cho, K. S. Hong, and J. Y. Kim, ACS Nano 7, 1027 (2013).

[15] H. R. Liu, J. H. Yang, H. J. Xiang, X. G. Gong, and S. H. Wei, Appl. Phys. Lett. 102, 112109 (2013).

[16] T. N. Stanislavchuk, A. A. Sirenko, A. P. Litvinchek, X. Luo, and S. –W. Cheong, J. Appl. Phys. 112, 044108 (2012).

[17] H. J. Kim, U. Kim, T. H. Kim, J. Kim, H. M. Kim, B. -G. Jeon, W. -J. Lee, H. S. Mun, K. T. Hong, J. Yu, K. Char, and K. H. Kim, Phys. Rev. B 86, 165205 (2012).

[18] L. Xie and J. Zhu, J. Am. Ceram. Soc., 95, 3597 (2012).

[19] E. Moreira, J. M. Henriques, D. L. Azevedo, E. W. S. Caetano, V. N. Freire, and E. L. Albuquerque, J. Solid State Chem. **184,** 921 (2011).

[20] K. Balamurugan, N. H. Kumar, J. A. Chelvane, and P. N. Santhosh, Physica B **407**, 2519 (2012).

[21] E. Moreira, J. M. Henriques, D. L. Azevedo, E. W. S. Caetano, V. N. Freire, and E. L. Albuquerque, J. Solid State Chem. **187**, 186 (2012).

[22] Bog G. Kim, J. Y. Cho, and S. W. Cheong, J. Solid State Chem. **197**, 134 (2013).

[23] A. M. Glazer, Acta Crystallogr. **28**, 3384 (1972).

[24] P. M. Woodward, Acta Crystallogr. **53**, 32 (1997).

[25] P. M. Woodward, Acta Crystallogr. **53**, 44 (1997).

[26] C. J. Howard and H. T. Stokes, Acta Crystallogr. **54**, 782 (1998).

[27] E. Bousquet, M. Dawber, M. Stucki, C. Lichtensteiger, P. Hermet, S. Gariglio, J. M. Triscone, and P. Ghosez, Nature **452**, 732 (2008).

[28] N. A. Benedek and C. J. Fennie, Phys. Rev. Lett. **106**, 107204 (2011).

[29] J. M. Rondinelli and C. J. Fennie, Adv. Mater. **24**, 1961 (2012).

[30] V. Gopalan and D. B. Litvin, Nature Materials **10**, 376 (2011).

[31] J. P. Perdew and A. Zunger, Phys. Rev. B **23**, 5048 (1981).

[32] G. Kresse and J. Furthmuller, Phys. Rev. B **54**, 11169 (1996).

[33] G. Kresse and D. Joubert, Phys. Rev. B **59**, 1758 (1999).

[34] H. J. Monkhorst and J. D. Pack, Phys. Rev. B **13**, 5188 (1976).

[35] A. Togo, F. Oba, and I. Tanaka, Phys. Rev. B **78**, 134106 (2008).

[36] R.D. King-Smith and D. Vanderbilt, Phys. Rev. B **47**, 1651 (1993); D. Vanderbilt and R.D. King-Smith, ibid. **48**, 4442 (1993); R. Resta, Rev. Mod. Phys. **66**, 899 (1994)

[37] ISOTROPY http://stokes.byu.edu/isotropy.html.

[38] D. Orobengoa, C. Capillas, M. I. Aroyo and J. M. Perez-Mato, J. Appl. Cryst. **42**, 820 (2009).

[39] F. D. Murnaghan, PNAS **30**, 244 (1944).

[40] See supplementary material at http://link.aps.org/supplemental/...

[41] After submitting the manuscript, we were aware of similar calculation has been done by other group ( http://arxiv.org/abs/1205.5526 ).

**Figure Captions**

Figure 1. The detailed structure of perovskite with various octahedral tilting, (a) $Pm\bar{3}m$ cubic without any tilting, (b) $I4/mcm$ tetragonal with $a^0a^0c^-$ tilting, (c) $Pnma$ orthorhombic with $a^-a^-c^+$ tilting. (d) Total energy vs Volume diagram of $CaSnO_3$ with three different symmetries. (e) Total Energy of three different phases of $BaSnO_3$, $SrSnO_3$, and $CaSnO_3$. (f) Tilting angels ($a^-$ and $c^+$) in orthorhombic phase of $BaSnO_3$, $SrSnO_3$, and $CaSnO_3$.

Figure 2. The detailed structure of superlattice $ASnO_3/BSnO_3$ (a) without octahedral tilting ($P4/mmm$ tetragonal) and (b) with $a^-a^-c^+$ octahedral tilting ($Pmc2_1$ orthorhombic). Total energy vs volume of tetragonal and orthorhombic phase of (c) $BaSnO_3/SrSnO_3$, (d) $BaSnO_3/CaSnO_3$, and (e) $SrSnO_3/CaSnO_3$ superlattice. (f) Tilting angels ($a^-$ and $c^+$) in the orthorhombic phase of three different superlattices.

Figure 3. Phonon dispersion curve of $P4/mmm$ phase of (a) $BaSnO_3/SrSnO_3$, (b) $BaSnO_3/CaSnO_3$, and (c) $SrSnO_3/CaSnO_3$ superlattice. Note that negative frequency is actually unstable imaginary frequency. Phonon dispersion curve of $Pmc2_1$ orthorhombic phase of (d) $BaSnO_3/SrSnO_3$, (e) $BaSnO_3/CaSnO_3$, and (f) $SrSnO_3/CaSnO_3$ superlattice.

Figure 4. Band structure and atom projected density of state in $Pmc2_1$ orthorhombic phase of (a) $BaSnO_3/SrSnO_3$, (b) $BaSnO_3/CaSnO_3$, and (c) $SrSnO_3/CaSnO_3$ superlattice. (d) Spontaneous polarization of three superlattices.

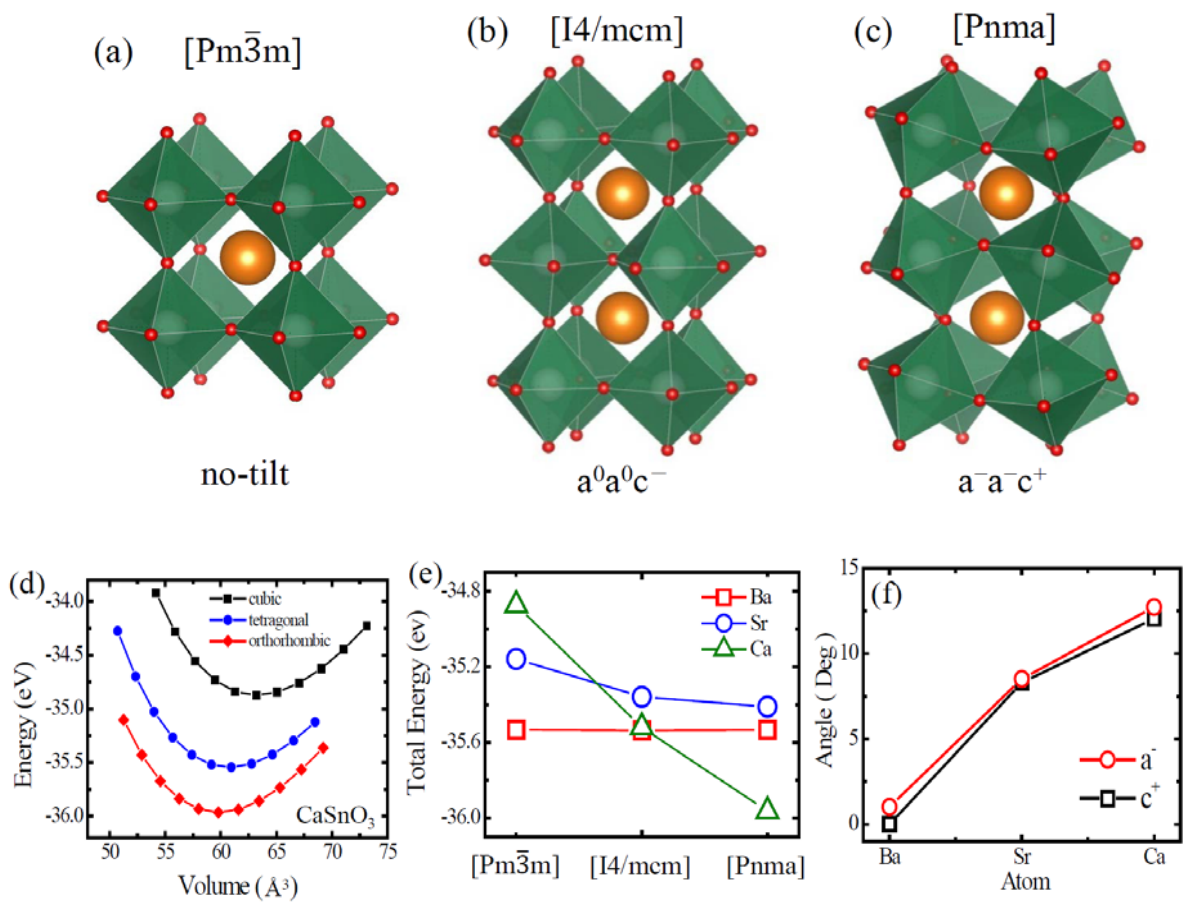

Figure 1 (Color Online) Sim, Cheong, and Kim

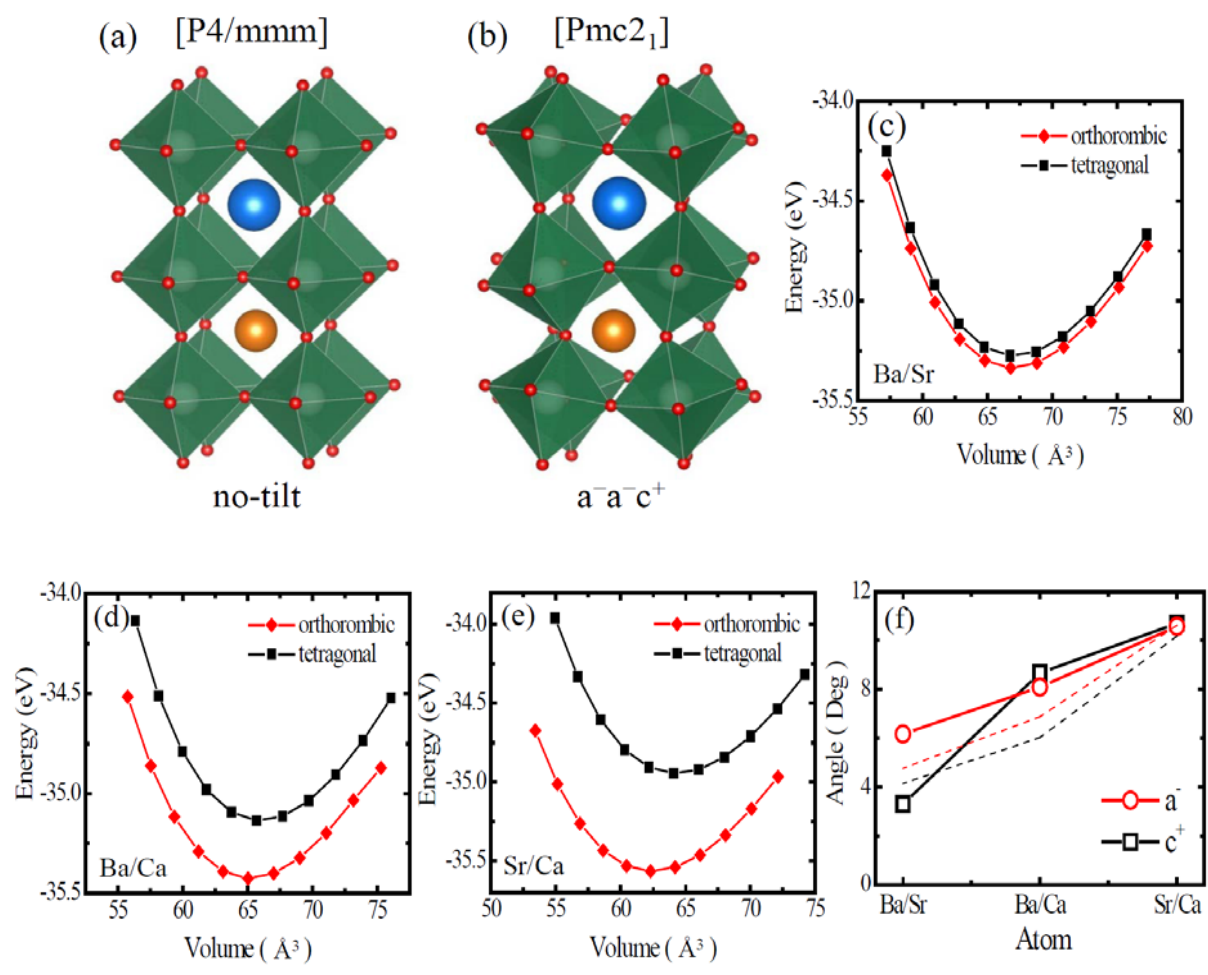

Figure 2 (Color Online) Sim, Cheong, and Kim

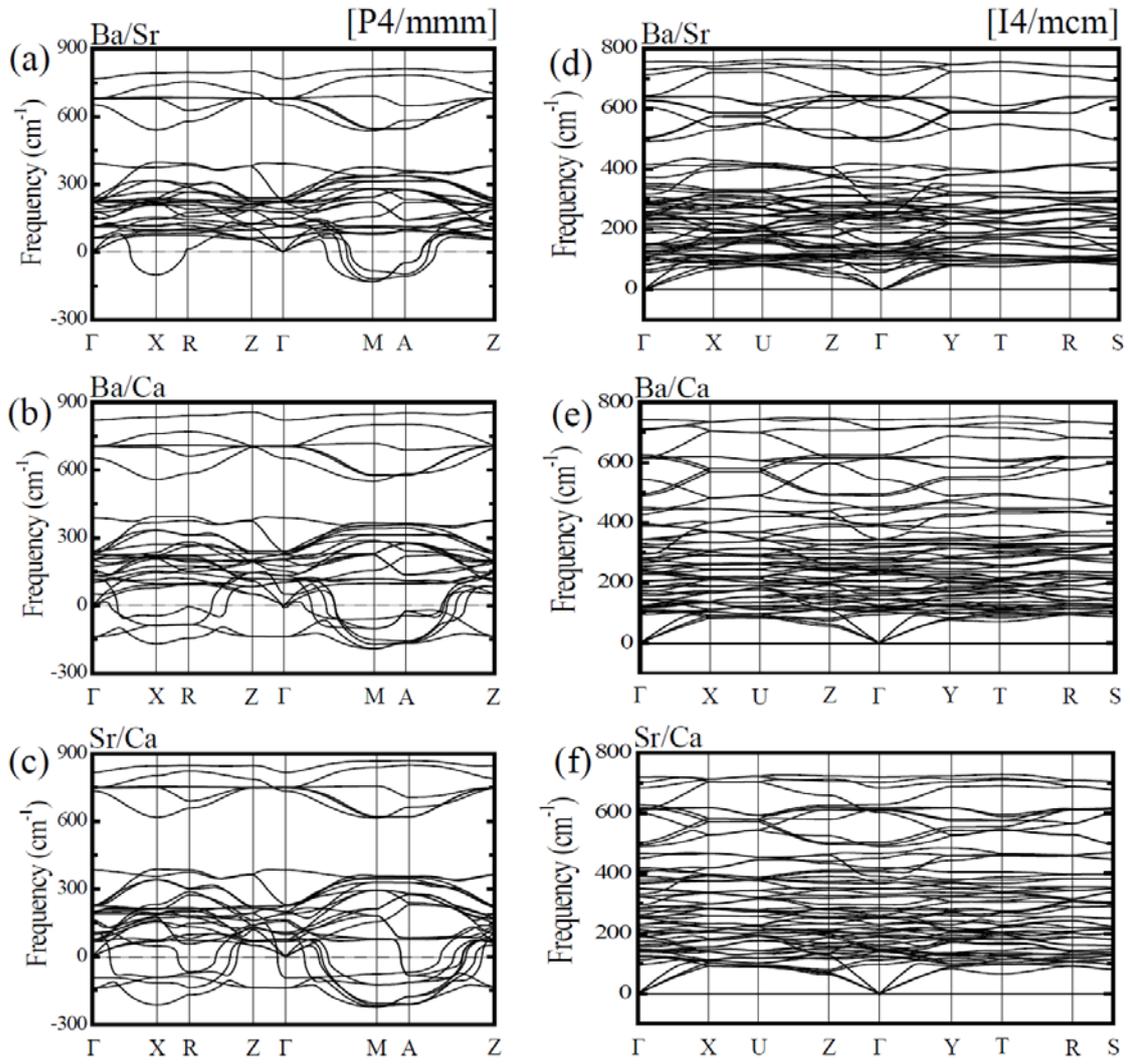

Figure 3 Sim, Cheong, and Kim

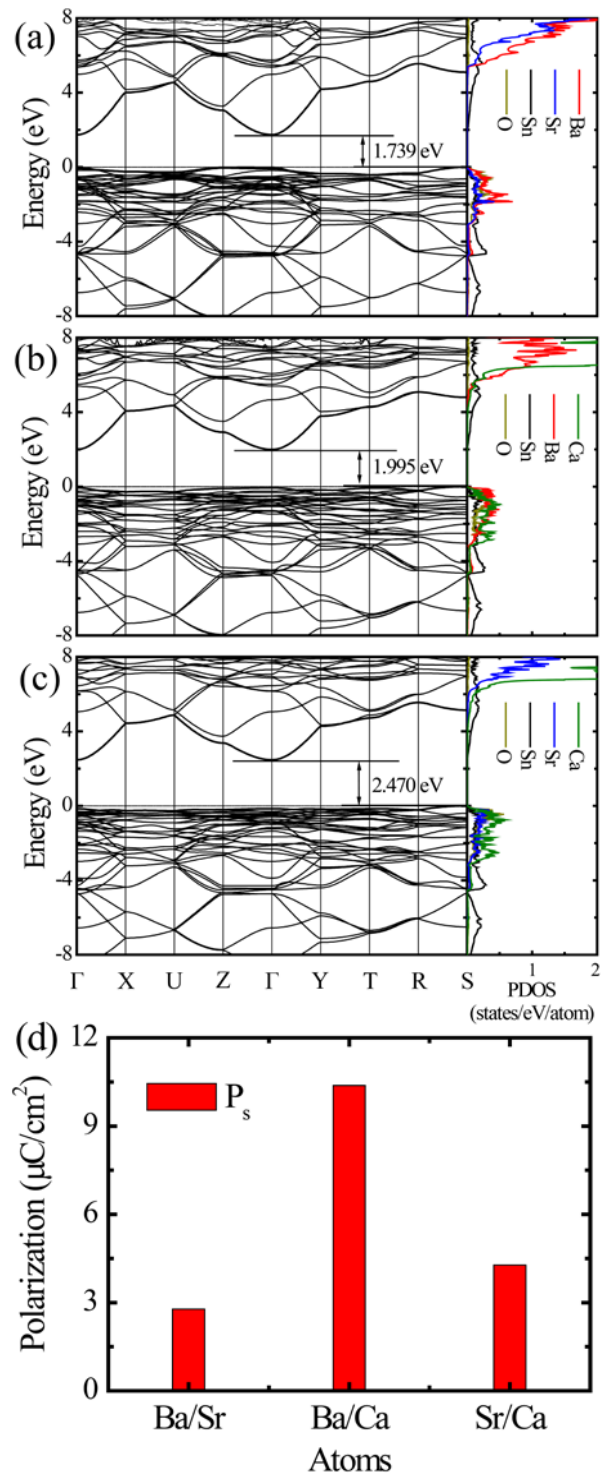

Figure 4 (Color Online) Sim, Cheong, and Kim

TABLE I. Structural and electronic properties of superlattices. The lattice constant, band gap energy, total energy, and polarization value are given for each space group (orthorhombic and tetragonal) of superlattices. Note that all orthorhombic phase have lower energy than tetragonal phase and spontaneous polarization is found in only orthorhombic phase.

| Atom | Space group | Lattice parameter (Å) | | | Band gap (eV) | Total energy ($ABO_3$ unit) | Polarization ($\mu C/m^2$) |
|---|---|---|---|---|---|---|---|
| | | a | B | c | | | |
| Ba/Sr | $Pmc2_1$ | 5.767 | 5.732 | 8.090 | 1.739 | -35.335 | 2.778 |
| | P4/mmm | 4.057 | 4.057 | 8.110 | 1.375 | -35.274 | 0.0 |
| Ba/Ca | $Pmc2_1$ | 5.651 | 5.710 | 8.062 | 1.995 | -35.424 | 10.379 |
| | P4/mmm | 4.037 | 4.037 | 8.064 | 1.551 | -35.135 | 0.0 |
| Sr/Ca | $Pmc2_1$ | 5.546 | 5.674 | 7.922 | 2.470 | -35.564 | 4.277 |
| | P4/mmm | 4.002 | 4.002 | 8.002 | 1.885 | -34.944 | 0.0 |

TABLE II. Amplimode result of superlattices. K-vector, character, and amplitude of each mode are given for three superlattices in orthorhombic phases. $\Gamma_{5-}$ mode is ferroelectric mode, $M_{5-}$, $M_{3+}$, and $M_{2+}$ modes are octahedral tilting modes.

| k-vector | Character | Amplitude (mode) | | |
|---|---|---|---|---|
| | | Ba/Sr | Ba/Ca | Sr/Ca |
| (0,0,0) | Ferroelectric | 0.1517 ($\Gamma_{5-}$) | 0.5431 ($\Gamma_{5-}$) | 0.6682 ($\Gamma_{5-}$) |
| (1/2,1/2,0) | Tilting ($a^-a^-c^0$) | 0.9174 ($M_{5-}$) | 1.2080 ($M_{5-}$) | 1.5147 ($M_{5-}$) |
| (1/2,1/2,0) | Tilting ($a^0a^0c^+$) | 0.3013 ($M_{3+}$) | 0.8667 ($M_{2+}$) | 1.0549 ($M_{2+}$) |

**Supplementary information**

Hyunsu Sim[1], S. W. Cheong[2], and Bog G. Kim[1,*]

[1] Department of Physics, Pusan National University, Pusan, 609-735, South Korea
[2] Rutgers Center for Emergent Materials and Department of Physics and Astronomy, Rutgers University, Piscataway, NJ 08854, USA.

Some of detailed calculation results are summarized here. Other results can be easily obtainable using the parameters presented here. Input and Output files for some calculations can be provided upon e-mail request (boggikim@pusan.ac.kr).

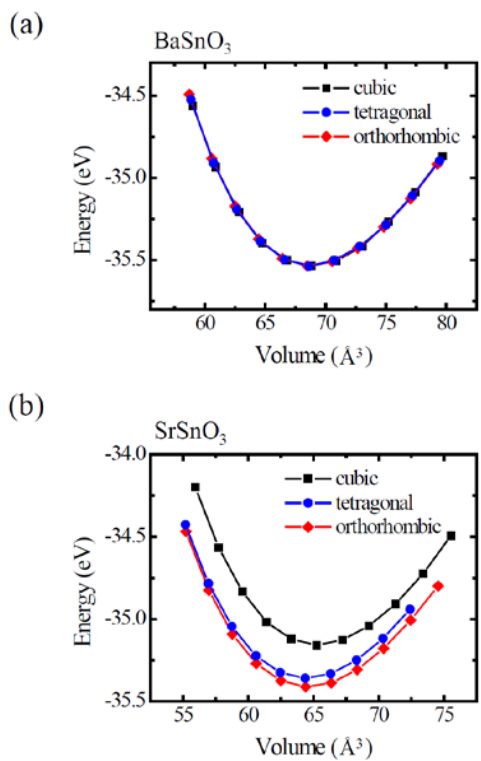

Figure S1. Total Energy as a function of unit cell volume for (a) $BaSnO_3$ and (b) $SrSnO_3$ in cubic, tetragonal, and orthorhombic phase.

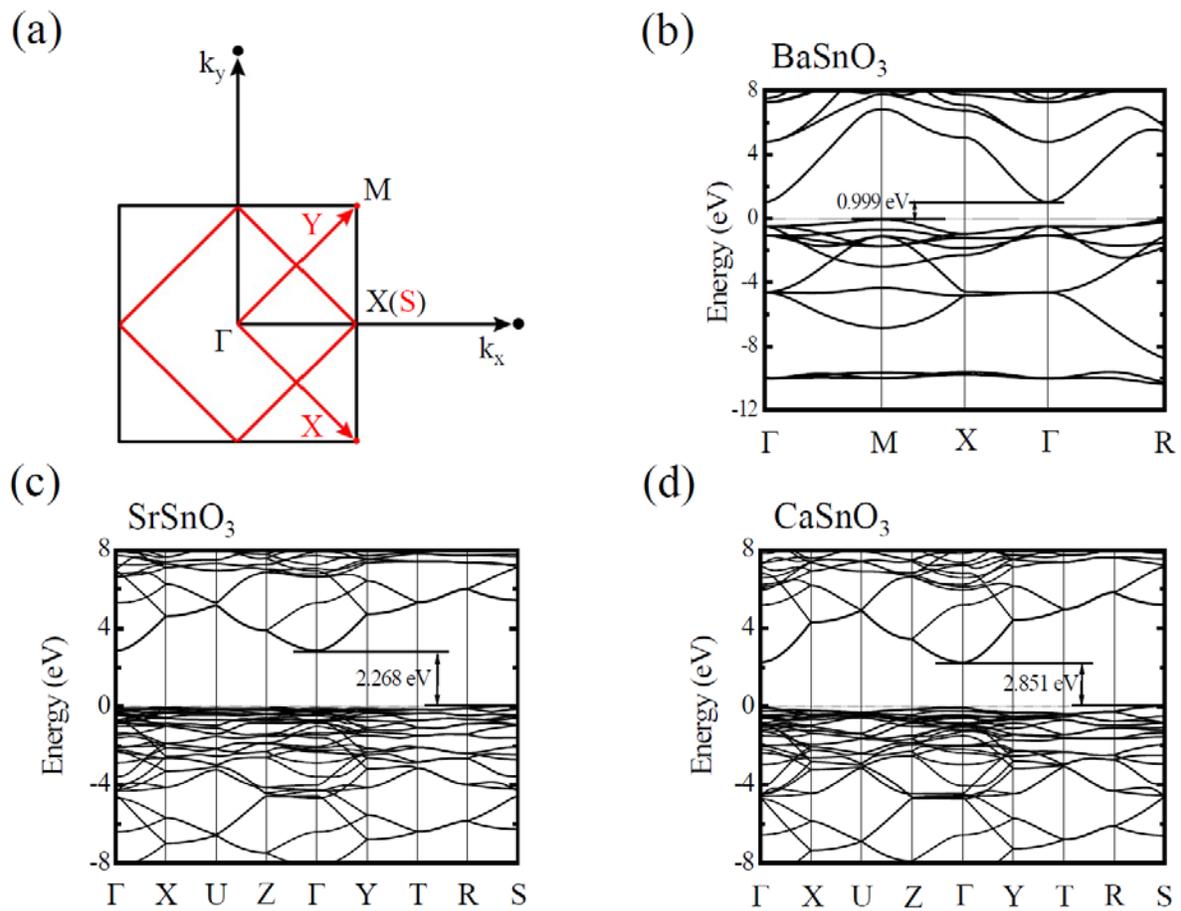

Figure S2. (a) Brillouin Zone of cubic phase and orthorhombic phase. Band Structure along specific Brillouin zone for (b) cubic $BaSnO_3$, (c) orthorhombic $SrSnO_3$, and (d) orthorhombic $CaSnO_3$.

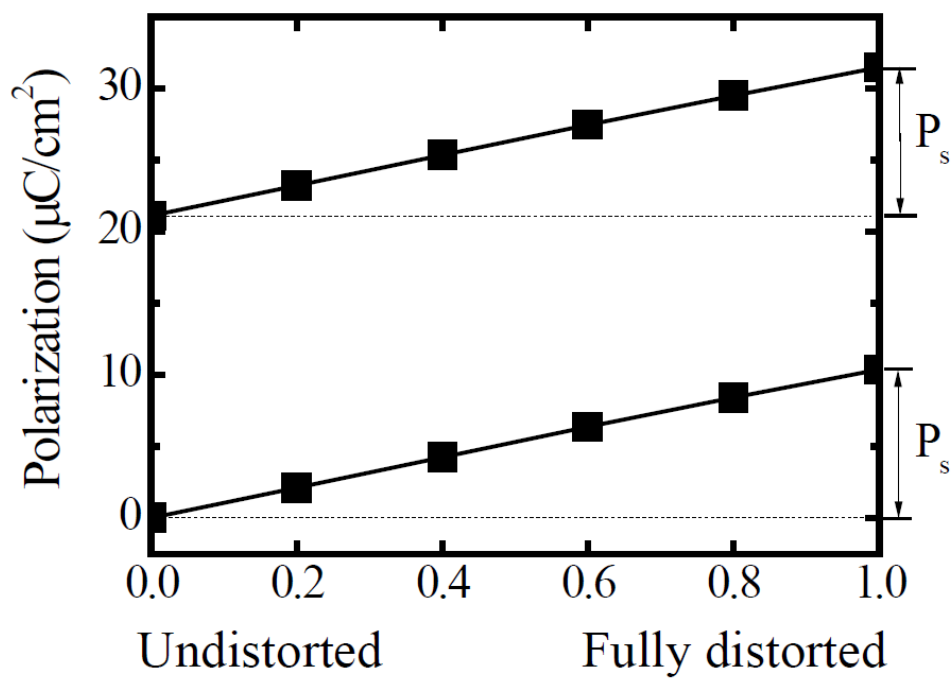

Figure S3. Berry phase polarization calculation of $BaSnO_3$/$CaSnO_3$ bicolor superlattice. Undistorted phase is tetragonal phase and fully distorted phase is orthorhombic ferroelectric phase. Polarization calculations have been done using different dipole center option in VASP to analyze polarization quantum and spontaneous polarization ($P_s$).

TABLE SI. Wyckoff positions of atoms in tetragonal and orthorhombic phase of superlattices

| Structure | Space group | Atom | Site | x | y | z |
|---|---|---|---|---|---|---|
| Ba/Sr | Pmc2$_1$ | Sr | 2a | 0.750 | 0.254 | 0.000 |
| | | Ba | 2b | 0.745 | 0.274 | 0.500 |
| | | Sn | 4c | 0.250 | 0.267 | 0.246 |
| | | O | 2a | 0.315 | 0.278 | 0.000 |
| | | O | 2b | 0.207 | 0.268 | 0.500 |
| | | O | 4c | 0.513 | 0.031 | 0.270 |
| | | O | 4c | 0.013 | 0.004 | 0.212 |
| Ba/Sr | P4/mmm | Sr | 1c | 0.500 | 0.500 | 0.000 |
| | | Ba | 1d | 0.500 | 0.500 | 0.500 |
| | | Sn | 2g | 0.000 | 0.000 | 0.253 |
| | | O | 4i | 0.000 | 0.500 | 0.257 |
| | | O | 1a | 0.000 | 0.000 | 0.000 |
| | | O | 1b | 0.000 | 0.000 | 0.500 |
| Ba/Ca | Pmc2$_1$ | Ca | 2a | 0.750 | 0.217 | 0.000 |
| | | Ba | 2b | 0.742 | 0.291 | 0.500 |
| | | Sn | 4c | 0.251 | 0.265 | 0.240 |
| | | O | 2a | 0.345 | 0.305 | 0.000 |
| | | O | 2b | 0.204 | 0.268 | 0.500 |
| | | O | 4c | 0.539 | 0.055 | 0.270 |
| | | O | 4c | 0.038 | 0.979 | 0.195 |
| Ba/Ca | P4/mmm | Ca | 1c | 0.500 | 0.500 | 0.000 |
| | | Ba | 1d | 0.500 | 0.500 | 0.500 |
| | | Sn | 2g | 0.000 | 0.000 | 0.245 |
| | | O | 4i | 0.000 | 0.500 | 0.238 |
| | | O | 1a | 0.000 | 0.000 | 0.000 |
| | | O | 1b | 0.000 | 0.000 | 0.500 |
| Sr/Ca | Pmc2$_1$ | Ca | 2a | 0.759 | 0.214 | 0.000 |
| | | Sr | 2b | 0.737 | 0.311 | 0.500 |
| | | Sn | 4c | 0.251 | 0.266 | 0.245 |
| | | O | 2a | 0.356 | 0.309 | 0.000 |
| | | O | 2b | 0.170 | 0.248 | 0.500 |
| | | O | 4c | 0.548 | 0.062 | 0.291 |
| | | O | 4c | 0.049 | 0.970 | 0.193 |
| Sr/Ca | P4/mmm | Ca | 1c | 0.500 | 0.500 | 0.000 |
| | | Sr | 1d | 0.500 | 0.500 | 0.500 |
| | | Sn | 2g | 0.000 | 0.000 | 0.252 |
| | | O | 4i | 0.000 | 0.500 | 0.255 |
| | | O | 1a | 0.000 | 0.000 | 0.000 |
| | | O | 1b | 0.000 | 0.000 | 0.500 |

TABLE SII. Atomic displacement of normal modes with given symmetry in the superlattices.

| Ba/Sr | | Reference structure | | | GM5- | | |
|---|---|---|---|---|---|---|---|
| Lattice constant | | 5.7673 | 5.7322 | 8.0902 | Ferroelectric | | |
| Atom | wyckoff | x | y | z | dx | dy | dz |
| Sr | 2a | 0.2500 | 0.7500 | 0.0000 | 0.0000 | -0.0939 | 0.0000 |
| Ba | 2b | 0.2500 | 0.7500 | 0.5000 | 0.0000 | 0.0434 | 0.0000 |
| Sn | 4c | 0.7500 | 0.7500 | 0.2458 | 0.0000 | -0.0046 | 0.0000 |
| O | 2a | 0.7500 | 0.7500 | 0.0000 | 0.0000 | 0.0668 | 0.0000 |
| O | 2b | 0.7500 | 0.7500 | 0.5000 | 0.0000 | -0.0012 | 0.0000 |
| O | 4c | 0.0000 | 0.5000 | 0.2413 | -0.0006 | -0.0015 | 0.0000 |
| O | 4c | 0.5000 | 0.0000 | 0.2413 | -0.0006 | -0.0015 | 0.0000 |
| | | M2+ | | | M5- | | |
| | | Tilting ($a^0a^0c^+$) | | | Tilting ($a^-a^-c^0$) | | |
| Atom | wyckoff | dx | dy | dz | dx | dy | dz |
| Sr | 2a | 0.0000 | 0.0000 | 0.0000 | 0.0002 | 0.0000 | 0.0000 |
| Ba | 2b | 0.0000 | 0.0000 | 0.0000 | 0.0053 | 0.0000 | 0.0000 |
| Sn | 4c | 0.0000 | 0.0000 | 0.0000 | 0.0001 | 0.0000 | 0.0000 |
| O | 2a | 0.0000 | 0.0000 | 0.0000 | -0.0708 | 0.0000 | 0.0000 |
| O | 2b | 0.0000 | 0.0000 | 0.0000 | 0.0471 | 0.0000 | 0.0000 |
| O | 4c | -0.0435 | -0.0435 | 0.0000 | 0.0000 | 0.0000 | -0.0314 |
| O | 4c | 0.0435 | 0.0435 | 0.0000 | 0.0000 | 0.0000 | 0.0314 |

| Ba/Ca | | Reference structure | | | GM5- | | |
|---|---|---|---|---|---|---|---|
| Lattice constant | | 5.6514 | 5.7099 | 8.0622 | Ferroelectric | | |
| Atom | wyckoff | x | y | z | dx | dy | dz |
| Ca | 2a | 0.7500 | 0.7500 | 0.0000 | 0.0000 | -0.0934 | 0.0000 |
| Ba | 2b | 0.7500 | 0.7500 | 0.5000 | 0.0000 | 0.0419 | 0.0000 |
| Sn | 4c | 0.2500 | 0.7500 | 0.2405 | 0.0000 | -0.0053 | 0.0000 |
| O | 2a | 0.2500 | 0.7500 | 0.0000 | 0.0000 | 0.0692 | 0.0000 |
| O | 2b | 0.2500 | 0.7500 | 0.5000 | 0.0000 | -0.0001 | 0.0000 |
| O | 4c | 0.0000 | 0.0000 | 0.2323 | 0.0007 | -0.0017 | 0.0000 |
| O | 4c | 0.5000 | 0.5000 | 0.2323 | 0.0007 | -0.0017 | 0.0000 |

|       |         | M2+       |           |        | M5−       |        |         |
|       |         | Tilting ($a^0a^0c^+$) | | | Tilting ($a^-a^-c^0$) | | |
| Atom  | wyckoff | dx        | dy        | dz     | dx        | dy     | dz      |
|-------|---------|-----------|-----------|--------|-----------|--------|---------|
| Ca    | 2a      | 0.0000    | 0.0000    | 0.0000 | 0.0000    | 0.0000 | 0.0000  |
| Ba    | 2b      | 0.0000    | 0.0000    | 0.0000 | −0.0070   | 0.0000 | 0.0000  |
| Sn    | 4c      | 0.0000    | 0.0000    | 0.0000 | 0.0005    | 0.0000 | 0.0000  |
| O     | 2a      | 0.0000    | 0.0000    | 0.0000 | 0.0787    | 0.0000 | 0.0000  |
| O     | 2b      | 0.0000    | 0.0000    | 0.0000 | −0.0384   | 0.0000 | 0.0000  |
| O     | 4c      | −0.0440   | −0.0440   | 0.0000 | 0.0000    | 0.0000 | −0.0312 |
| O     | 4c      | 0.0440    | 0.0440    | 0.0000 | 0.0000    | 0.0000 | 0.0312  |

| Sr/Ca  |         | Reference structure | | | GM5− | | |
| Lattice constant | | 5.5458 | 5.6740 | 7.9224 | Ferroelectric | | |
| Atom  | wyckoff | x       | y       | z      | dx        | dy      | dz     |
|-------|---------|---------|---------|--------|-----------|---------|--------|
| Ca    | 2a      | 0.7500  | 0.7500  | 0.0000 | 0.0000    | −0.0806 | 0.0000 |
| Sr    | 2b      | 0.7500  | 0.7500  | 0.5000 | 0.0000    | 0.0649  | 0.0000 |
| Sn    | 4c      | 0.2500  | 0.7500  | 0.2448 | 0.0000    | −0.0027 | 0.0000 |
| O     | 2a      | 0.2500  | 0.7500  | 0.0000 | 0.0000    | 0.0624  | 0.0000 |
| O     | 2b      | 0.2500  | 0.7500  | 0.5000 | 0.0000    | −0.0298 | 0.0000 |
| O     | 4c      | 0.0000  | 0.0000  | 0.2417 | −0.0008   | −0.0029 | 0.0000 |
| O     | 4c      | 0.5000  | 0.5000  | 0.2417 | −0.0008   | −0.0029 | 0.0000 |
|       |         | M2+     |         |        | M5−       |         |        |
|       |         | Tilting ($a^0a^0c^+$) | | | Tilting ($a^-a^-c^0$) | | |
| Atom  | wyckoff | dx      | dy      | dz     | dx        | dy      | dz     |
| Ca    | 2a      | 0.0000  | 0.0000  | 0.0000 | 0.0063    | 0.0000  | 0.0000 |
| Sr    | 2b      | 0.0000  | 0.0000  | 0.0000 | −0.0088   | 0.0000  | 0.0000 |
| Sn    | 4c      | 0.0000  | 0.0000  | 0.0000 | 0.0005    | 0.0000  | 0.0000 |
| O     | 2a      | 0.0000  | 0.0000  | 0.0000 | 0.0698    | 0.0000  | 0.0000 |
| O     | 2b      | 0.0000  | 0.0000  | 0.0000 | −0.0526   | 0.0000  | 0.0000 |
| O     | 4c      | −0.0446 | −0.0446 | 0.0000 | 0.0000    | 0.0000  | −0.0323 |
| O     | 4c      | 0.0446  | 0.0446  | 0.0000 | 0.0000    | 0.0000  | 0.0323  |